\documentclass[11pt,twoside]{article}
\usepackage{asp2010}

\resetcounters

\begin{document}

\title{Early Results from the Wisconsin H-Alpha Mapper \\ Southern Sky Survey}
\author{L.M.~Haffner$^1$, R.J.~Reynolds$^1$, G.J.~Madsen$^2$, A.S.~Hill$^1$, K.A.~Barger$^1$, K.P.~Jaehnig$^1$, E.J.~Mierkiewicz$^1$, J.W.~Percival$^1$, N.~Chopra$^1$
\affil{$^1$Department of Astronomy, University of Wisconsin---Madison,\\475 North Charter Street, Madison, WI 53706}
\affil{$^2$Sydney Institute for Astronomy, School of Physics A29,\\The University of Sydney, NSW 2006, Australia}}

\begin{abstract}
After a successful eleven-year campaign at Kitt Peak, we moved the Wisconsin H-Alpha Mapper (WHAM) to Cerro Tololo in early 2009. Here we present some of the early data after a few months under southern skies. These maps begin to complete the first all-sky, kinematic survey of the diffuse H$\alpha$ emission from the Milky Way. Much of this emission arises from the Warm Ionized Medium (WIM), a significant component of the ISM that extends a few kiloparsecs above the Galactic disk. While this first look at the data focuses on the H$\alpha$ survey, WHAM is also capable of observing many other optical emission lines, revealing fascinating trends in the temperature and ionization state of the WIM. Our ongoing studies of the physical conditions of diffuse ionized gas will continue from the southern hemisphere following the H$\alpha$ survey. In addition, future observations will cover the full velocity range of the Magellanic Stream, Bridge, and Clouds to trace the ionized gas associated with these neighboring systems.
\end{abstract}

\section{Introduction}

Although first detected through the absorption of diffuse, low-frequency radio synchrotron radiation \citep{HoyEll63}, the Warm Ionized Medium (WIM) has been studied primarily through optical emission lines over the past few decades. These observations have shown the WIM to be a pervasive component of the ISM with temperatures ranging 8,000 K to 12,000 K; ionization fractions (H$^+$/H) typically $>$ 90\%; a scale height near the Sun and within the Perseus Arm of $\sim$1 kpc; and typical local electron densities of 0.01 cm$^{-3}$  to 0.1 cm$^{-3}$ \citep{HafDetBec09}. Although denser, ``classical'' H \textsc{ii} regions contribute a comparable fraction of the total Galactic H$\alpha$ luminosity, the WIM contains roughly 90\% of the ionized mass of the Milky Way.

To fully map the extent and kinematics of the WIM, we designed and built a dedicated instrument that excels in observations of faint, diffuse optical emission. The Wisconsin H-Alpha Mapper (WHAM) consists of a steerable siderostat with a 0.6-m primary lens coupled to a 15-cm, dual-etalon Fabry-Perot spectrometer. This optical configuration (R $\sim$ 30,000) delivers a spatially integrated spectrum from a one-degree beam on the sky covering 200 km/s with 12 km/s spectral resolution. Optimized for H$\alpha$, a 30-second exposure per pointing provides sensitivity to emission with I $\sim$ 0.1~R (EM $\sim$ 0.2 pc cm$^{-6}$), allowing us to survey the observable sky in about two years.

\section{Survey Observations}

WHAM survey data is obtained in ``blocks'' of approximately 30--50 one-degree pointings organized into a roughly $7\deg \times 7\deg$ grid. Trading full sampling for full sky coverage in a tractable span of time ($\sim$ two years), the grid spacing is $\Delta b = 0.85\deg, \Delta \ell = 0.98 \times \cos b$. For each pointing, WHAM exposes for 30 s with the spectrometer configured to cover a spectral region approximately $-100$ km/s $\leq v_\mathrm{LSR} \leq+100$ km/s around H$\alpha$. Geocoronal H$\alpha$ emission distinct in all but the brightest Galactic regions provides a consistent, precise velocity reference calibration. Intensities are calibrated by frequent observations of standard nebular targets, which not only track instrumental performance, but also measure atmospheric transmission from night to night.

Apart from the geocorona, faint ($I < 0.1$ R) atmospheric emission lines are present throughout the 200 km/s band. In most directions above $|b| = 20\deg$, accurate removal of these lines is essential to recover Galactic intensities and trace structures in the faint background. \citet{WHAMNSS} details the procedure we followed for the Northern Sky Survey (NSS) to clean and calibrate that data. The new data presented here from CTIO have only the geocorona subtracted, but the Galactic emission is also predominantly bright. The imprint of the atmospheric lines can still be seen at higher velocities in faint regions; for example, the elevated background in the lower-left of the first channel maps of Figure~\ref{fig:gum} is due to lines that have not been removed from the southern data yet. The general pattern of these faint lines appears to be similar at CTIO compared to KPNO, but we will be carefully analyzing the sky spectra from the faintest blocks before attempting a complete subtraction of the atmosphere from the southern survey.

Over the next year, we will complete a full survey of the southern sky ($\delta < +30\deg$), with an emphasis on the newly visible portion. Overlap with the NSS will anchor consistent calibration between the surveys. In addition, we now have access to the complex, extended gas systems associated with the Magellanic Clouds, including the Bridge, Stream, and Leading Arm. Although much of the velocity extent of these structures will not be covered by the survey proper, we will lead a campaign to provide the most sensitive measures---and hopefully maps---of ionized gas associated with these halo systems. We will also return to multi-wavelength observations of diffuse gas and continue our investigation of the physical conditions of the WIM.  

\section{The Gum Nebula}

\articlefigure{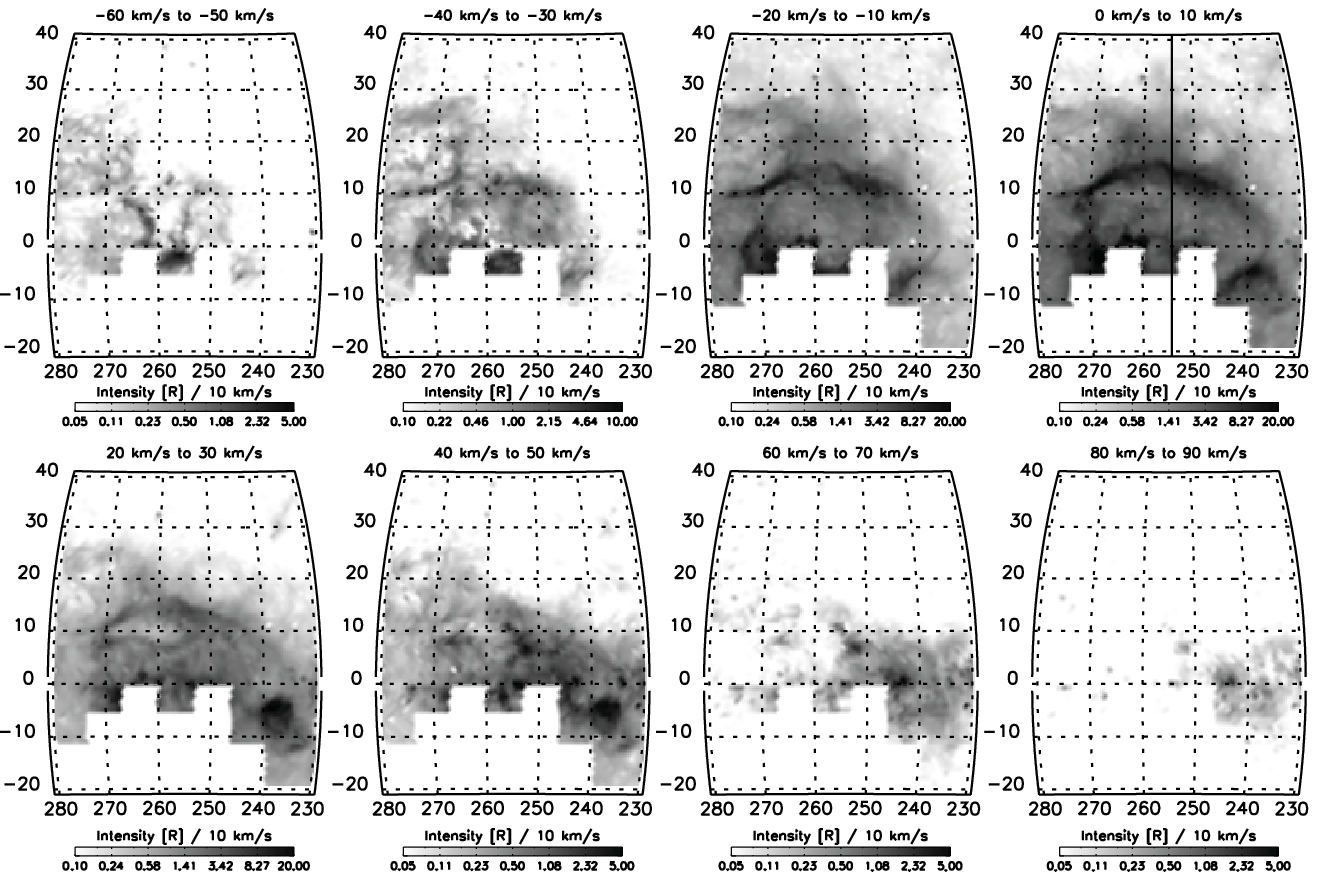}{fig:gum}{\textbf{Channel Maps of the Gum Nebula.} Each map shows the integrated intensity of a 10 km/s channel. The intensity scale is logarithmic, and the scale endpoints change gradually from map to map to maximize feature visibility. The black line in the 0 to $+10$ km/s map denotes the longitude (254\deg) of the latitude-velocity diagram shown in Figure~\ref{fig:bv}.}

One of the first regions WHAM observed after installation at CTIO is the northern half of one of the largest ionized structures on the sky, the Gum Nebula. Relatively close ($\sim$500 pc) and first recognized over 50 years ago \citep{Gum52}, its extent and nature have been debated over the decades since due to its large size and diffuse emission \citep*{Rey76,Rey76a,ChaSiv83,SahSah93,WoeGayOtr01,ChoBha09}. With its sensitivity and spectral capabilities, WHAM is ideally suited to provide a comprehensive, multi-wavelength view of the ionized emission from the nebula. Such an investigation is one of the primary objectives of our efforts after the completion of the H$\alpha$ survey.

Figure~\ref{fig:gum} shows a series of H$\alpha$ channel maps from the Gum region using data from both the northern and southern surveys. Although targeted spectral observations have been reported previously \citep{SahSah93,Rey76}, Figure~\ref{fig:gum} presents the first kinematic map of the ionized gas from a portion of the nebula. When the survey is complete, we will be able to explore the gas dynamics of the entire nebula and separate emission components coincident along the line of sight. For example, the most prominent feature in the $+20$ km/s to $+30$ km/s map of Figure~\ref{fig:gum} centered at $\ell = 238\deg,\ b = -5\deg$ (Sh 2-310) is associated with the more distant ($\sim$1500 pc) open cluster NGC 2362.

\articlefigure[scale=0.75]{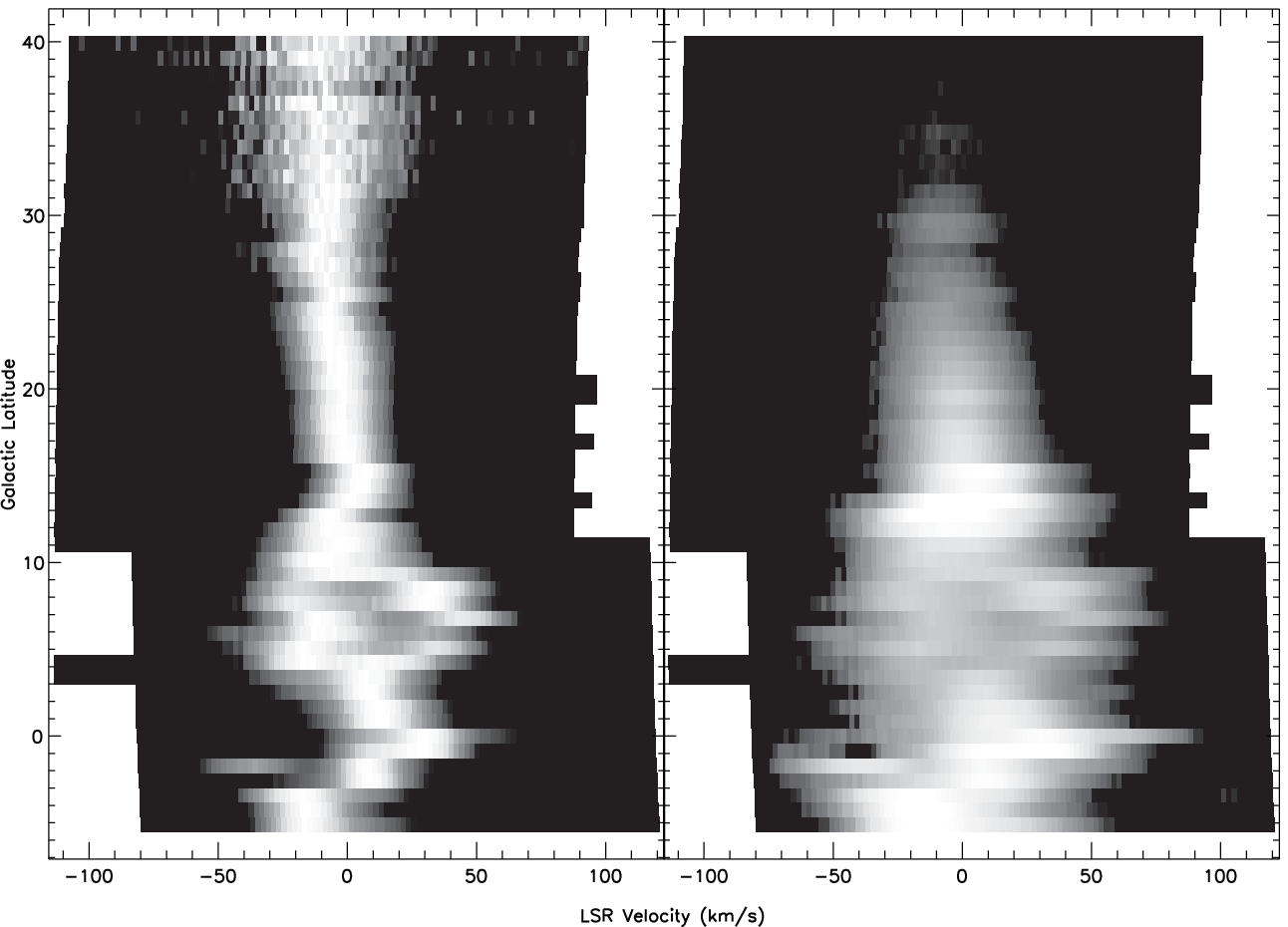}{fig:bv}{\textbf{Latitude-Velocity Diagrams at \boldmath $\ell = 254\deg$.} This 2\deg-wide slice displays each spectrum as an image at its respective latitude. The left panel allows each spectrum to use the full intensity scale so that components stand out regardless of absolute intensity. The right panel enforces a global, logarithmic intensity scale that spans a factor of 250 from black to white, allowing the intensity distribution to be tracked. Note the double-peaked profiles around $b = 3\deg$ to 10\deg, which range $\pm20$ to 30 km/s from the systemic velocity of the nebular shell edge, 0 to $+10$ km/s.}

Full coverage of spectral observations also leads to a more comprehensive kinematic analysis, such as the latitude-velocity diagrams presented in Figure~\ref{fig:bv}. The black line in the 0 to $+10$ km/s channel of Figure~\ref{fig:gum} shows the location of the $b$-$v$ slice. A few interesting features immediately stand out. First, the highest latitude emission feature ($b \sim +15\deg$) commonly associated with the Gum---although not always, see \citet{WoeGayOtr01} for an alternative hypothesis---shows a velocity clearly different from the WIM emission above it. Continuing toward lower latitudes, although the emission is fainter, the component structure is considerably more complex, with distinct red and blue components prominent at $b = +3\deg$ to $+10\deg$. \citet{Rey76} conjectured that this behavior is due to an expansion of the nebula ($\sim$ 20 km/s), which is also supported in more recent OH observations of molecular gas in the region \citep{WoeGayOtr01}.

This sample of survey data toward the Gum Nebula highlights some of the unique features of WHAM. After the H$\alpha$ observations are completed, we will return to the region and followup with maps of other diagnostic lines such as H$\beta$, [N~\textsc{ii}] $\lambda6584$, [S~\textsc{ii}] $\lambda6716$, He~\textsc{i} $\lambda5876$, and [O~\textsc{iii}] $\lambda5007$ to trace the physical conditions throughout the region. Many locations will also be bright enough for deeper observations with fainter lines, such as [O~\textsc{i}] $\lambda6300$ and [N~\textsc{ii}] $\lambda5755$. Such a multiline analysis will provide the most comprehensive kinematic study of the optical emission from the Gum Nebula to date and help  explore the possible scenarios for its creation and current powering.

\acknowledgements The move from KPNO to CTIO as well as ongoing WHAM operations are supported by NSF award AST-0607512. WHAM was built with the help of the University of Wisconsin Graduate School, Physical Sciences Lab, and Space Astronomy Lab. Sam Gabelt and Don Michalski provided much needed expertise during the 2008 upgrade in Wisconsin. The relocation south was smooth substantially due to the excellent staff at KPNO and CTIO. Without them we would not be able to remotely observe under such fine skies, collecting more than a decade of fantastic data.

\bibliography{haffner2_l}

\end{document}